\documentclass{jfm}
\usepackage{graphicx}
\usepackage{mathrsfs}
\usepackage{amsmath}
\usepackage{amssymb}
\usepackage{bm}
\usepackage{color}
\usepackage{epstopdf, epsfig}
\newcommand{\dd}{\mathrm{d}}
\newcommand{\DD}{\mathrm{D}}
\newcommand{\Pe}{\mathrm{Pe}}
\newcommand{\St}{\mathrm{St}}
\newcommand{\andand}{\,\,\,\text{and}\,\,\,}
\newcommand{\with}{\,\,\,\text{with}\,\,\,}

\newcommand{\comma}{,\,\,\,}

\def\So{\mathscr{S}}

\newcommand{\specialnumber}[1]{%
	\def\tagform@##1{\maketag@@@{(\ignorespaces##1\unskip\@@italiccorr#1)}}}

\shorttitle{Dispersion of inertial particles in cellular flows}
\shortauthor{A. Renaud and J. Vanneste}

\title{Dispersion of inertial particles in cellular flows in the small-Stokes, large-P\'eclet regime}

\author{Antoine Renaud\aff{1}
  \corresp{\email{antoine.renaud@ed.ac.uk}} \and
  Jacques Vanneste\aff{1}}

\affiliation{\aff{1}School  of  Mathematics  and  Maxwell  Institute  for  Mathematical  Sciences,  University  of  Edinburgh, King’s Buildings, Edinburgh EH9 3FD, United Kingdom}

\begin{document}

\maketitle

\begin{abstract}
We investigate the transport of inertial particles by cellular flows when advection dominates over inertia and diffusion, that is, for Stokes and  P\'eclet numbers satisfying $\St \ll 1$ and $\Pe \gg 1$. Starting from the Maxey--Riley model, we consider the distinguished scaling $\St \, \Pe = O(1)$ and derive an effective Brownian dynamics approximating the full Langevin dynamics. We then apply homogenisation and matched-asymptotics techniques to obtain an explicit expression for the effective diffusivity $\overline{D}$ characterising long-time dispersion. This expression quantifies how $\overline{D}$, proportional to $\Pe^{-1/2}$ when inertia is neglected, increases for particles heavier than the fluid and decreases for lighter particles. In particular, when $\St \gg \Pe^{-1}$, we find that $\overline{D}$ is proportional to $\St^{1/2}/(\log ( \St \, \Pe))^{1/2}$ for heavy particles and exponentially small in $\St \, \Pe$ for light particles. We verify our asymptotic predictions against numerical simulations of the particle dynamics. 
\end{abstract}


\section{Introduction\label{sec:1}}

Over long time scales, the dispersion of particles and passive scalars in periodic incompressible flows is asymptotically a pure diffusive process, with an effective diffusivity tensor which can be computed, e.g.\ using the method of homogenisation \citep{Majda1999, Pavliotis2005}.  The flows that have attracted most attention in this context are shear flows and cellular flows because of the remarkably different dependence of their effective diffusivities on the P\'eclet number and the availability of closed-form results. The  cellular flow -- on which this paper focuses -- is the two-dimensional flow with the stream function
\begin{equation}\label{eq:CellularFlow}
    \psi(x,y)= U a \sin (x/a)\sin (y/a),
\end{equation}
where $U$ is the maximum flow speed and $2\pi a$ is the cell period. This flow consists of a doubly periodic array of cells containing four vortices, with the fluid rotating alternatively clockwise and anti-clockwise in each quarter-cell. A classic result, due to \citet{Childress1979}, \citet{Shrainman1987}, \citet{Rosenbluth1987} and \citet{Soward1987}, gives the asymptotic form of the (isotropic) effective diffusivity for non-inertial particles when advection dominates over diffusion. Using $a$ and $U/a$ as reference length and time, it reads 
\begin{equation}\label{eq:Diff_passive}
    \overline{D} \sim 2\So\Pe^{-1/2} \quad \textrm{for} \ \ \Pe\gg 1, 
\end{equation}
where $\Pe=Ua/D$ is the P\'eclet number and $D$ is the diffusion coefficient. The constant $\So\approx0.5327\cdots$ was determined by \cite{Soward1987} using Wiener--Hopf techniques. (See also the mathematical literature, e.g.\ \citet{Heinze2003,Novikov2005} for rigorous bounds.)

\begin{figure}
    \centering
    \includegraphics[width=0.7\linewidth]{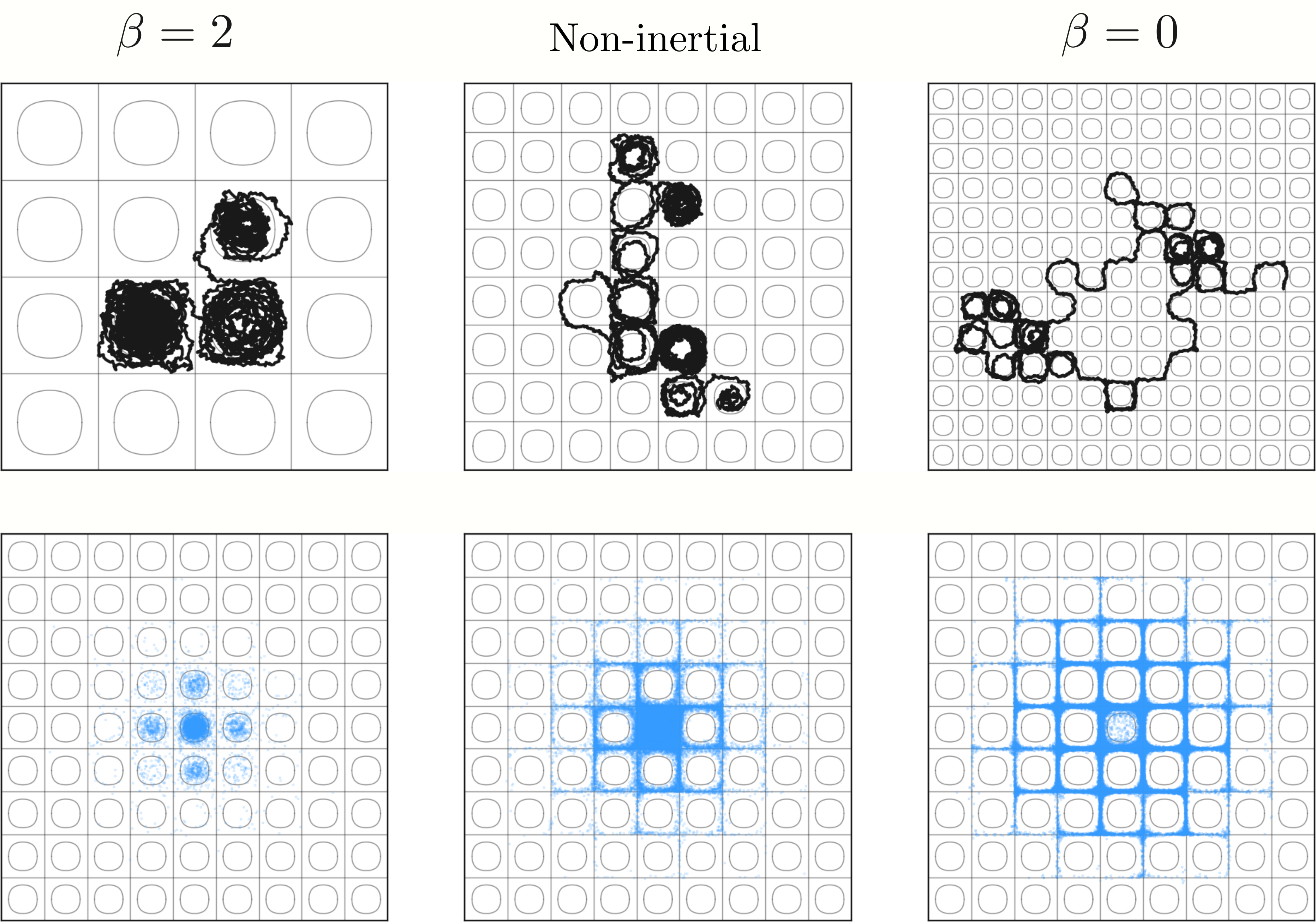}
    \caption{Single particle trajectory $\bm{X}(t)$ for $0\leq t\leq 500$ (top row) and position at $t=30$ of $10^5$ particles initially  distributed  uniformly in a quarter-cell (bottom row), with thin lines indicating the streamlines $\psi=0,\pm0.5$. The non-inertial results (centre) are obtained by integrating \eqref{eq:PassivScal_Langevin} with $\Pe=1000$; the results for inertial particles are obtained by integrating \eqref{eq:Langevin} with $\Pe=1000$, $\St=0.1$ and $\beta=2$ (left) and $\beta=0$ (right).}
    \label{fig:Sample}
\end{figure}

In this paper, we investigate how the dispersion is affected by small-but-finite particle inertia. Numerical simulations \citep[e.g.][]{Pavliotis2006, Pavliotis2009} show a strong enhancement of the effective diffusivity compared to the non-inertial case when the particles are denser than fluid. This is because inertia expels heavy particles away from high-vorticity regions, towards the cellular flow's separatrices, enhancing transport between cells and hence large-scale dispersion. Conversely, particles less dense than the fluid tend to accumulate in the (high-vorticity) cell centres, leading to a much smaller effective diffusivity. Our aim is to quantify this by generalising \eqref{eq:Diff_passive} to inertial particles. 

The dynamics of non-inertial particles is Brownian, governed by the  nondimensional equation
\begin{equation}\label{eq:PassivScal_Langevin}
 \dd\bm{X}=\bm{u}\,\dd t+\left(2/\Pe\right)^{1/2}\dd\bm{W},   
\end{equation}
where $\bm{X}$ is the particle position, $\bm{W}$ a two-dimensional Wiener process, and $\bm{u}=(\partial_y\psi,-\partial_x\psi)$ is the fluid velocity. We model the dynamics of inertial particles by the Langevin equation based on the Maxey--Riley (\citeyear{MaxeyRiley1983}) model,
\begin{subequations}\label{eq:Langevin}
\begin{align}
            \dd\bm{X}&=\bm{V}\dd t,\\
            \St\, \dd\bm{V}&=\left(\bm{u}\left(\bm{X},t\right)-\bm{V}\right)\dd t+\St\beta \, \DD_t\bm{u}\dd t+(2/\Pe)^{1/2}\dd \bm{W}, \label{eq:LangevinMom}
\end{align}    
\end{subequations}
where $\bm{V}$ is the particle velocity, $\DD_t=\partial_t+\bm{u}\cdot\nabla$ is the material derivative along the fluid velocity $\bm{u}$, $\St=\tau U/a$ is the Stokes number, with $\tau=r_\mathrm{p}^2/(9\beta\nu)$ the Stokes timescale, $\beta=3\rho_\mathrm{f}/(2\rho_\mathrm{p}+\rho_\mathrm{f})\in[0,3]$ is the density parameter with $\rho_\mathrm{f}$ and $\rho_\mathrm{p}$ the mass density of the fluid and particle, $r_\mathrm{p}$ is the particle radius, and $\nu$ is the kinematic viscosity of the fluid. The momentum equation \eqref{eq:LangevinMom} includes Stokes drag (first term on the right-hand-side) and Auton's added-mass force (second term, see \cite{Autom1988}) but neglects the Boussinesq--Basset force and Faxen correction as well as gravity. 

Typical examples of single trajectories, obtained by solving \eqref{eq:Langevin} for light ($\beta > 1$) and heavy ($\beta < 1$) particles and \eqref{eq:PassivScal_Langevin} for non-inertial particles, are shown in the top row of Figure \ref{fig:Sample}. The early-time dispersion is illustrated by the bottom row which shows the location at a fixed time $t$ of an ensemble of particles initially concentrated within a single quarter-cell. The figure demonstrates the expected enhancement of dispersion for heavy particles and inhibition for light particles.

\begin{figure}
    \centering
    \includegraphics[width=\linewidth]{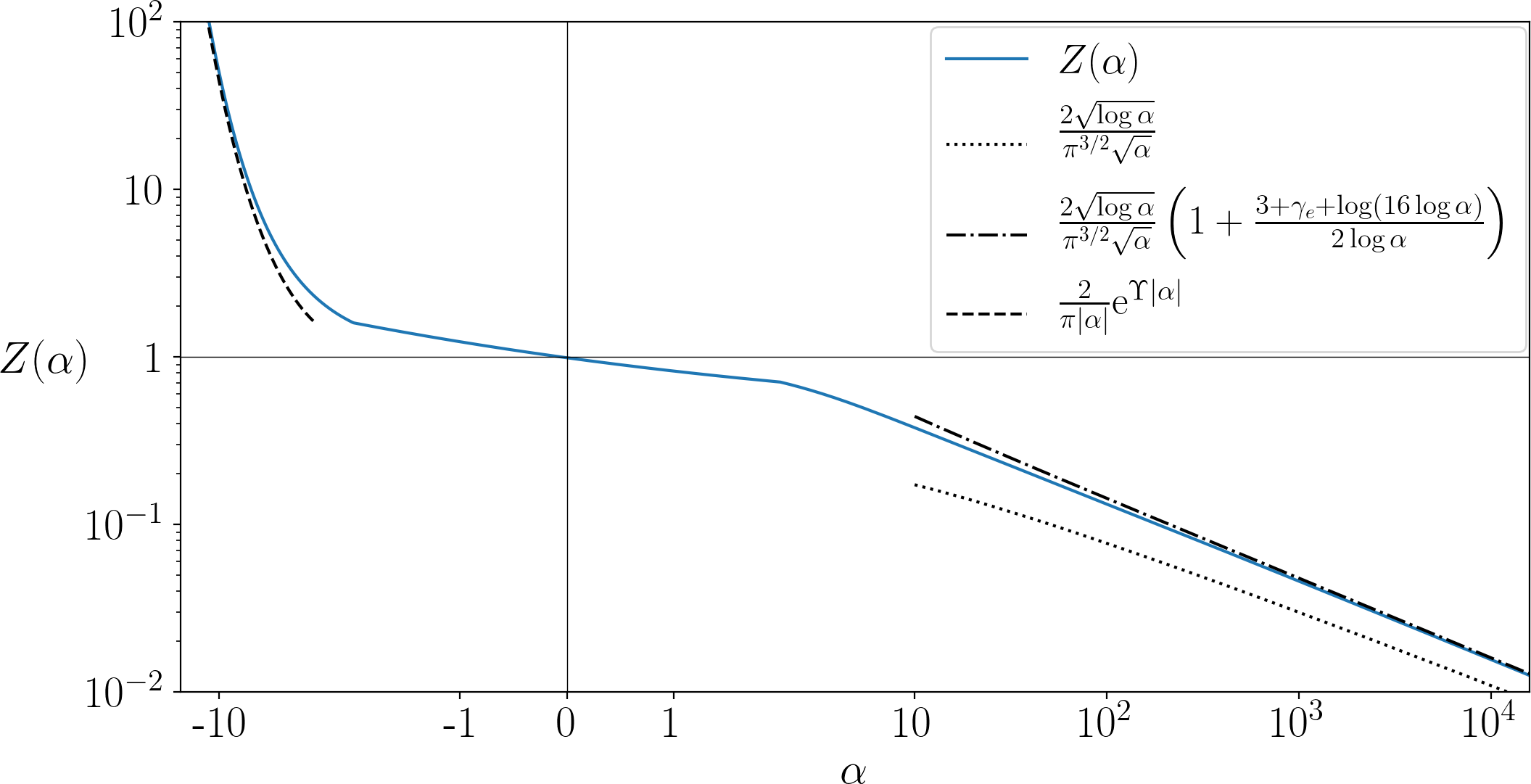}
    \caption{Log--log plot of $Z(\alpha)$ defined by \eqref{eq:Za}. Asymptotic expressions  for $\alpha \to \pm \infty$ obtained in appendix \ref{ssec:App:asympZ} are shown as dashed and dot-dashed lines ($\gamma_\mathrm{E} = 0.577\cdots$ is the Euler--Mascheroni constant and $\Upsilon=0.655\cdots$ is defined in \eqref{eq:Za_Arheynius}).}
    \label{fig:Za}
\end{figure}

The Brownian dynamics \eqref{eq:PassivScal_Langevin} in which the velocity rather than the acceleration is white in time is recovered from the Langevin dynamics \eqref{eq:Langevin} when inertial effects are negligible compared to diffusion, that is, for $\St \ll \Pe^{-1}\ll 1$. 
We consider the more general, distinguished regime  $\St \ll 1$, $\Pe \gg 1$ with $\St \, \Pe =O(1)$, where both inertial and diffusive effects enter the problem at the same order. {We emphasise that  analysing this distinguished regime makes it possible to treat the entire range of relative values of $\St$ and $\Pe$ at once, including the limiting cases $\St \ll \Pe^{-1}\ll 1$ and $\Pe^{-1} \ll  \St \ll 1$.}
Our main result is an asymptotic formula for the effective diffusivity in this regime:
\begin{equation}\label{eq:Effective_Diff_MainResult}
\overline{D} \sim \frac{2{\So} \Pe^{-1/2}}{Z(\alpha)} \quad \textrm{for} \ \St, \, \Pe^{-1} \ll 1,\ \St \, \Pe =O(1).
\end{equation}
Here,  
\begin{equation}
    \alpha = \St\,\Pe \, (1-\beta)
    \label{eq:alpha}
\end{equation}
and $Z(\alpha)$ is a function given explicitly in terms of elliptic integrals in \eqref{eq:Za} below and shown in Figure \ref{fig:Za}.  Comparing \eqref{eq:Effective_Diff_MainResult} with its non-inertial counterpart \eqref{eq:Diff_passive} shows that $Z(\alpha)$ completely captures the effect of inertia. Since it is monotonically decreasing and satisfies $Z(0)=1$, inertia is confirmed to decrease $\overline{D}$ for light particles ($\alpha<0$) and to increase $\overline{D}$ for heavy particles ($\alpha>0$). 

To derive \eqref{eq:Effective_Diff_MainResult}, we first reduce the Langevin dynamics \eqref{eq:Langevin} to an effective Brownian dynamics which captures weak inertia (\S\ref{sec:2}). We then apply the method of homogenisation, formulate its cell problem in \S\ref{sec:3}, and solve it asymptotically  in \S\ref{sec:4}, recasting the results of  \citet{Childress1979}, \citet{Shrainman1987}, \citet{Rosenbluth1987} and \citet{Soward1987} in the language of homogenisation along the way. We discuss the results, derive simplified versions of \eqref{eq:Effective_Diff_MainResult} valid when inertia dominates diffusion ($\St \, \Pe \gg 1$), and conclude in \S\ref{sec:5}. {The reader uninterested in the details of the computation can skip to \S \ref{sec:5}.}

\section{Effective Brownian dynamics\label{sec:2}}

In this section, we derive an effective Brownian dynamics capturing inertial effects in the distinguished regime $\St, \, \Pe^{-1} \ll 1$ with $\St \, \Pe = O(1) $ and $\beta=O(1)$. The derivation is similar to that carried out in \citet{Pavliotis2009} in the case $\beta=0$.
It is convenient to introduce a small parameter $\epsilon \ll 1$ such that $\Pe^{-1}=\epsilon^2$, let $\St=\gamma\epsilon^2$ with $\gamma=O(1)$,
and define the rescaled relative velocity
$\bm{P}=\gamma\epsilon\left(\bm{V}-\bm{u}\right)$ to rewrite \eqref{eq:Langevin} as
\begin{subequations} \label{eq:Langevin_P}
\begin{align}
            \dd\bm{X}&=\left(\bm{u}+\bm{P}/(\gamma\epsilon)\right)\dd t, \\            \dd\bm{P}&=-\left(\bm{P}/(\gamma\epsilon^2)+\gamma \epsilon(1-\beta)\DD_t\bm{u}-\left(\bm{P}\cdot\nabla\right)\bm{u}\right)\dd t+\sqrt{2} \, \dd \bm{W}.
\end{align}
\end{subequations}
We reduce the dynamics of \eqref{eq:Langevin_P} by considering the corresponding backward Kolmogorov equation, namely
\begin{equation}\label{eq:BKE}
    \partial_t g-\bm{u}\cdot\nabla_{\bm{x}}g=\left(\frac{1}{\gamma\epsilon}\bm{p}\cdot\nabla_{\bm{x}} -\left(\frac{1}{\gamma \epsilon^2}\bm{p}+\gamma \epsilon(1-\beta)\DD_t\bm{u}-\left(\bm{p}\cdot\nabla_{\bm{x}}\right)\bm{u}\right)\cdot\nabla_{\bm{p}}+\nabla_{\bm{p}}^2\right)g
\end{equation}
for {functions
\begin{equation}
g(\bm{x},\bm{p},t)=\mathbb{E}[f(\bm{X}(t),\bm{P}(t))|\bm{X}(0)=\bm{x},\bm{P}(0)=\bm{p}],
\end{equation}
where $\mathbb{E}$ denotes the expectation over the Wiener process $\bm{W}$ and $f$ is an arbitrary function \citep[see e.g.][]{Evans2013}.}
We now introduce the multiple time scales $t_n=\epsilon^n t$ with $n=0,1,\cdots$ and the expansion
 \begin{equation}\label{eq:g_expan}
     g\left(\bm{p},\bm{x},t\right)=g_0\left(\bm{x},t_0,t_1,t_2,\cdots\right)+\epsilon g_1\left(\bm{p},\bm{x},t_0,t_1,t_2,\cdots\right)+\cdots.
 \end{equation}
Substituting \eqref{eq:g_expan} into  \eqref{eq:BKE} yields, up to order $\epsilon^2$,
\begin{subequations}
\label{eq:gAllorders}
 \begin{align}
     \frac{\bm{p}}{\gamma }\cdot\left(\nabla_{\bm{x}}g_0-\nabla_{\bm p}g_1\right)&=0\label{eq:g_order-1},\\
     \frac{\bm{p}}{\gamma}\cdot\left(\nabla_{\bm{x}}g_1-\nabla_{\bm{p}} g_2\right)&=\left(\partial_{t_0}-\bm{u}\cdot\nabla_{\bm{x}}\right)g_0,\label{eq:g_order0}\\
     \frac{\bm{p}}{\gamma }\cdot\left(\nabla_{\bm{x}}g_2-\nabla_{\bm p}g_3\right)&=\left(\partial_{t_0}-\bm{u}\cdot\nabla_{\bm{x}}\right)g_1+\partial_{t_1}g_0-\nabla_{\bm{p}}^2g_1-\left(\bm{p}\cdot\nabla_{\bm{x}}\right)\bm{u}\cdot\nabla_{\bm{p}}g_1 ,\label{eq:g_order1}\\
    \frac{\bm{p}}{\gamma }\cdot\left(\nabla_{\bm{x}}g_3-\nabla_{\bm p}g_4\right)&=\left(\partial_{t_0}-\bm{u}\cdot\nabla_{\bm{x}}\right)g_2+\partial_{t_1}g_1-\nabla_{\bm{p}}^2g_2- \left(\bm{p}\cdot\nabla_{\bm{x}}\right)\bm{u}\cdot\nabla_{\bm{p}}g_2 \nonumber \\
     &+\partial_{t_2}g_0+\gamma (1-\beta)\DD_t\bm{u}\cdot\nabla_{\bm{x}}g_0.\label{eq:g_order2}
 \end{align}
 \end{subequations}
Using that $g_0$ is independent of $\bm{p}$, we can solve \eqref{eq:gAllorders} successively.  Solving \eqref{eq:g_order-1} then \eqref{eq:g_order0} yields 
\begin{subequations}
    \begin{align}
        g_1=\bm{p}\cdot\nabla_{\bm{x}}g_0,&\quad
        g_2=\frac{\bm{p}\cdot\nabla_{\bm{x}}}{2}g_1,\label{eq:g_12_Sol} \\
        \partial_{t_0}g_0&=\bm{u}\cdot\nabla_{\bm{x}}g_0\label{eq:g_t0_Sol}
    \end{align}
\end{subequations}
Introducing \eqref{eq:g_12_Sol}  into \eqref{eq:g_order1} and using \eqref{eq:g_t0_Sol} gives
\begin{equation}
    \frac{1}{\gamma}\left(\frac{(\bm{p}\cdot\nabla_{\bm{x}})^3}{2}g_0-\bm{p}\cdot\nabla_{\bm{p}}g_3\right)=\partial_{t_1}g_0,
\end{equation}
which is satisfied by $g_3={(\bm{p}\cdot\nabla_{\bm{x}})^3}g_0/{6}$ and
\begin{equation}
    \partial_{t_1}g_0=0.\label{eq:g_t1_Sol}
\end{equation}
Finally, introducing \eqref{eq:g_12_Sol} into \eqref{eq:g_order2}, using \eqref{eq:g_t0_Sol} and \eqref{eq:g_t1_Sol} and setting $\bm{p}=0$ yields
\begin{equation}\label{eq:g_t2_Sol}
    \partial_{t_2}g_0=-\gamma(1-\beta)\DD_t\bm{u}\cdot\nabla_{\bm{x}}g_0+\nabla_{\bm{x}}^2g_0.
\end{equation}

From  \eqref{eq:g_t0_Sol}, \eqref{eq:g_t1_Sol} and \eqref{eq:g_t2_Sol}, it is clear that advection by $\bm{u}$ is the dominant process while inertia and  diffusion arise at the same, lower order. We reconstitute the dynamics of $g_0$ capturing both all three effects by adding \eqref{eq:g_t0_Sol}, \eqref{eq:g_t1_Sol} and \eqref{eq:g_t2_Sol}, setting $\partial_{t_0}+\epsilon\partial_{t_1}+\epsilon^2\partial_{t_2}= \partial_t$ and using $\epsilon^2 = \Pe^{-1}$ and $\gamma \epsilon^2=\St$ to obtain the backward Kolmogorov equation
\begin{equation}\label{eq:BKE_Overdamped}
    \partial_t g_0=\left(\bm{u}-\St(1-\beta)\DD_t\bm{u}\right)\cdot \nabla g_0+\frac{1}{\Pe}\nabla^2 g_0
\end{equation}
for $g_0$. This corresponds to the effective Brownian (or overdamped Langevin) dynamics 
\begin{equation}\label{eq:Overdamped-Langevin}
    \dd\bm{X}=\left(\bm{u}-\St(1-\beta)\DD_t\bm{u}\right)\dd t+(2/\Pe)^{1/2}\dd\bm{W},
\end{equation}
which includes inertial correction through the term $\St(1-\beta)\DD_t\bm{u}$. In the absence of diffusion ($\Pe\to\infty$), \eqref{eq:Overdamped-Langevin} recovers the so-called first-order Eulerian approximation \cite[e.g.,][]{Ferry2001} or equivalently the slow manifold dynamics  discussed by \cite{Rubin1995} and \cite{Haller2008}.

According to \eqref{eq:Overdamped-Langevin}, inertial particles behave as though advected by the effective flow 
\begin{equation}\label{eq:EffectivFlow}
\bm{u}_{\mathrm e}=\bm{u}-\St(1-\beta)\DD_t\bm{u}.
\end{equation}
For incompressible fluids, $\bm{u}$ is divergence free but $\bm{u}_{\rm e}$ rarely is. 
It can be shown that 
\begin{equation}\label{eq:DivU}
    \nabla\cdot\bm{u}_{\rm e}=-\St(1-\beta)\left(\left\|\bm{S}\right\|^2-\left\|\bm{\Omega}\right\|^2\right),
\end{equation}
where $\bm{S}$ and $\bm{\Omega}$ are the local strain-rate and rotation-rate tensors of  $\bm{u}$. 
Thus particles denser than the fluid  ($\beta<1$) tend to accumulate ($\nabla\cdot\bm{u}_{\rm e}<0$) in low-vorticity and high-strain regions,  while  less-dense particles ($\beta>1$) tend to accumulate in high-vorticity and low-strain regions. A detailed analysis of this clustering effect without diffusion is given by \citet{Haller2010}. {Isodense particles ($\beta=1$) are unaffected by inertia to the order of accuracy of \eqref{eq:EffectivFlow}. Non-zero higher-order terms resulting from the coupling between inertia and noise appear when the expansion of the backward Kolmogorov equation is carried out to $O(\St^2)$; these involve derivatives of $g$ of order higher than 2, characterising a non-diffusive correction to the trajectories $\bm{X}(t)$. Note that differences between the behaviour of rigid isodense particles and the fluid they replace also arise from finite-size (Faxen) effects which we neglect.}

\section{Periodic homogenisation \label{sec:3}}

In this section, we apply the methodology of homogenisation to the Brownian dynamics \eqref{eq:Overdamped-Langevin}  to obtain an expression for the corresponding effective diffusivity. The derivation is standard {\citep[see e.g.][]{Vergassola1997,Pavliotis2005}} and recorded here for completeness and to set up notation. 

We rewrite the associated backward Kolmogorov equation \eqref{eq:BKE_Overdamped} as 
\begin{equation}\label{eq:BKE_U}
    \partial_tg=\bm{u}_{\rm e}\cdot\nabla g+\frac{1}{\Pe}\nabla^2g,
\end{equation}
where $\bm{u}_{\rm e}=\bm{u}-\St(1-\beta)\bm{u}\cdot\nabla\bm{u}$ is the effective velocity field  \eqref{eq:EffectivFlow} which is steady, periodic and divergent.
We introduce the small parameter $\delta\ll1$ along with the variables $\bm{X}=\delta\,\bm{x}$,   $t_1=\delta\,t$ and $t_2=\delta^2t$. We seek for a solution of \eqref{eq:BKE_U} in the form
\begin{equation}\label{eq:g_hom_expand}
    g(\bm{x},t)=g_0\left(\bm{X},t_1,t_2\right)+\delta g_1\left(\bm{x},\bm{X},t_1,t_2\right)+\delta^2g_2\left(\bm{x},\bm{X},t_1,t_2\right)\cdots\,.
\end{equation} 
where the $g_i, \, i=1,2,\cdots,$ are periodic functions of $\bm{x}$.
We introduce the expansion \eqref{eq:g_hom_expand} into \eqref{eq:BKE_U} and collect terms at each order in $\delta$. At orders $\delta$ and $\delta^2$, we find
\begin{subequations}
\begin{align}
    \partial_{t_1}g_0&=\mathcal{L}_1g_0+\mathcal{L}_0g_1,\label{eq:g_hom_order1}\\
    \partial_{t_2}g_0&=\mathcal{L}_0 g_2+\mathcal{L}_1 g_1+\mathcal{L}_2g_0\label{eq:g_hom_order2},
\end{align}
with 
\begin{equation}
    \mathcal{L}_0=\bm{u}_{\rm e}\cdot\nabla_{\bm{x}}+\frac{1}{\Pe}\nabla_{\bm{x}}^2\comma\mathcal{L}_1=\bm{u}_{\rm e}\cdot\nabla_{\bm{X}}+\frac{2}{\Pe}\nabla_{\bm{X}}\cdot\nabla_{\bm{x}}\andand\mathcal{L}_{2}=\frac{1}{\Pe}\nabla_{\bm{X}}^2.
\end{equation}
\end{subequations}
Let $\phi^{\dagger}\left(\bm{x}\right)$ be the invariant measure associated with $\mathcal{L}_0$, that is, the periodic solution of
\begin{equation}\label{eq:L0dag}
    \mathcal{L}_0^{\dagger}\phi^{\dagger}=-\bm{u}_{\rm e}\cdot\nabla_{\bm{x}}\phi^{\dagger}+\frac{1}{\Pe}\nabla_{\bm{x}}^2\phi^{\dagger}=0
\end{equation}
normalised such that $\langle\phi^{\dagger}\rangle=1$, with $\langle\cdot\rangle$ denoting  spatial average over the periodic cell. Multiplying \eqref{eq:g_hom_order1} by $\phi^{\dagger}$ and averaging yields
\begin{equation}
    \partial_{t_1}g_0=\bm{c}\cdot\nabla_{\bm{X}}g_0,\label{eq:hom_order1_av}
\end{equation}
where 
\begin{equation} \label{eq:c}
    \bm{c}=\left\langle\phi^{\dagger}\bm{u}_{\rm e}\right\rangle
\end{equation}
is the effective drift velocity vector.

Using \eqref{eq:hom_order1_av} and \eqref{eq:g_hom_order1}, $g_1$ is found to satisfy
\begin{equation}\label{eq:g_1_hom_celprob}
    \mathcal{L}_0g_1+\left(\bm{u}_{\rm e}-\bm{c}\right)\cdot\nabla_{\bm{X}}g_0=0.
\end{equation}
The solution takes the form $g_1=\bm{\varphi}\cdot\nabla_{\bm{X}}g_0$, where $\bm{\varphi}$ obeys the cell problem
\begin{equation}\label{eq:hom_varphi_cellprob}
    \mathcal{L}_0\bm{\varphi}+\bm{u}_{\rm e}-\bm{c}=0
\end{equation}
with periodic boundary conditions.
Substituting this solution into \eqref{eq:g_hom_order2}, multiplying by $\phi^{\dagger}$ and averaging yields the effective equation
\begin{equation}\label{eq:hom_order2_av_bis}
    \partial_{t_2}g_0=\nabla_{\bm{X}}\cdot\left(\DD_{\rm e}\cdot\nabla_{\bm{X}}\right)g_0,
\end{equation}
where 
\begin{equation}\label{eq:Diff_-1}
    \DD_{\rm e}=\left\langle\phi^{\dagger}\left(\frac{1}{\Pe}\left(\mathbb{I}+2\nabla_{\bm{x}}\otimes\bm{\varphi}\right)+\left(\bm{u}_{\rm e}-\bm{c}\right)\otimes\bm{\varphi}\right)\right\rangle,
\end{equation}
is the effective diffusivity tensor with $\mathbb{I}$ denoting the identity tensor and $\otimes$ the tensorial product. Finally, we gather \eqref{eq:hom_order1_av} and \eqref{eq:hom_order2_av_bis}, set $\delta\partial_{t_1}+\delta^2\partial_{t_2}=\partial_t$ and $\delta\nabla_{\bm{X}} = \nabla$ to obtain the effective backward Kolmogorov equation  
\begin{equation}\label{eq:Diffusion_equation}
    \partial_t g_0=\bm{c}\cdot\nabla g_0+\nabla\cdot\left(\DD_{\rm e}\cdot\nabla\right)g_0.
\end{equation}
This corresponds to a purely diffusive process, characterised by the  effective drift velocity $\bm{c}$ and diffusivity tensor $\DD_{\rm e}$, which approximates the dynamics \eqref{eq:Overdamped-Langevin} over long time scales.

Computing $\bm{c}$ and $\DD_{\rm e}$, requires solving both \eqref{eq:L0dag} for $\phi^\dagger$ and \eqref{eq:hom_varphi_cellprob} for $\bm{\varphi}$. In the non-inertial case, $\nabla\cdot\bm{u}_{\rm e}=0$ and the invariant measure $\phi^{\dagger}$ is simply a constant. For inertial particles, $\phi^{\dagger}$ is non-trivial and reflects the clustering caused by the divergence of $\bm{u}_{\rm e}$. 

We remark that we obtained \eqref{eq:c} and \eqref{eq:Diff_-1} by first deriving the effective Brownian dynamics \eqref{eq:Overdamped-Langevin} then applying homogenisation. Alternatively, we could have first applied homogenisation to the Langevin dynamics \eqref{eq:Langevin}, then exploited the asymptotic parameters to simplify the cell problems. {We have checked that this alternative route yields the same cell problem; in other words, the limits $\St \to 0$ and $\delta \to 0$ commute.}  \cite{Pavliotis2005} and \cite{afonso2012} {take $\delta \to 0$ first followed by $\St \to 0$,
but they do not consider the small-$\Pe$ asymptotics.}
 We note that the cell problems for the Langevin dynamics are difficult to solve as they are defined within domains including unbounded velocity spaces.  The sampling of trajectories is also delicate for the Langevin dynamics when $\St \ll 1$ since it requires exceedingly small time steps, and we make use of the effective Brownian dynamics for this purpose also below.
 
In the next section, we solve the cell problems \eqref{eq:L0dag} and \eqref{eq:hom_varphi_cellprob}  asymptotically in the distinguished regime $\St=O(\Pe^{-1})\ll1$ and compute the leading order expression of the effective diffusivity \eqref{eq:Diff_-1} for the cellular flow. 

\section{Effective diffusivity\label{sec:4}}

\begin{figure}
    \centering
    \includegraphics[width=\linewidth]{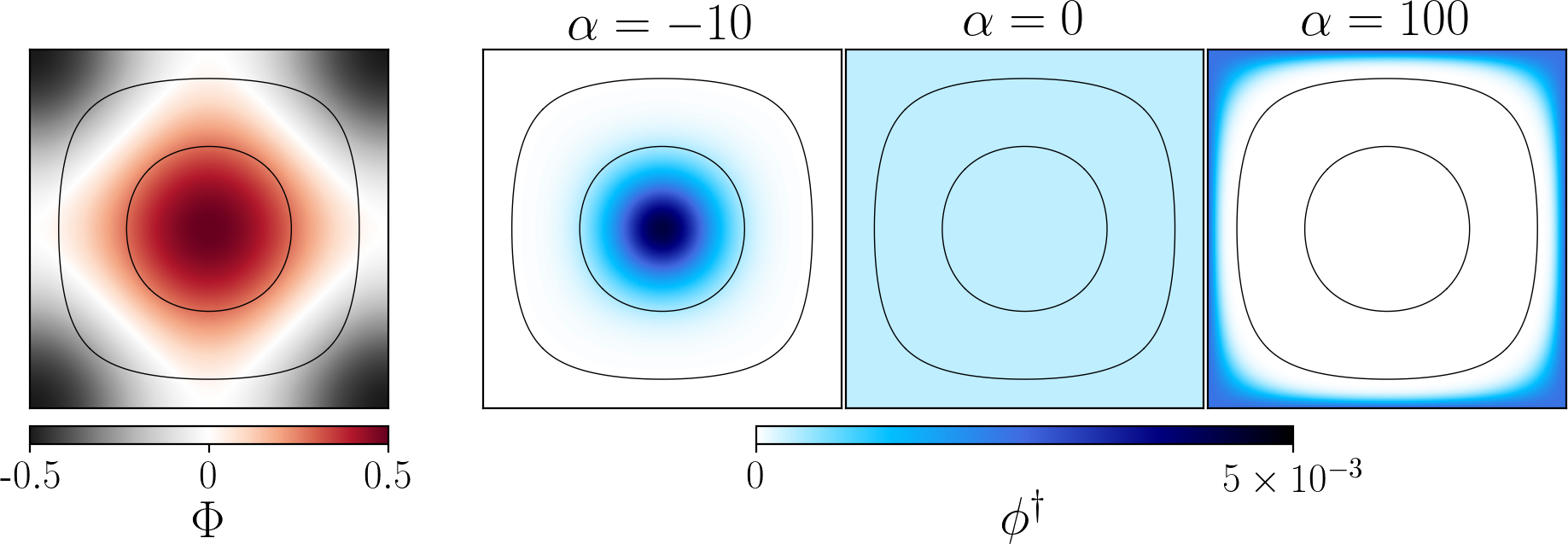}
    \caption{Potential $\Phi$ defined in \eqref{eq:CellularFlow_c} (left panel) and  invariant measure $\phi^{\dagger}$ for $\alpha=-10$, $0$ and $100$ computed from \eqref{eq:phidag_FInal} (right panels). Both fields are $\pi$-periodic and shown here in a quarter-cell. The thin dark lines represent the streamlines $|\psi|=0$, $0.25$ and $0.75$.}
    \label{fig:phidag}
\end{figure}

In non-dimensional variables, the effective velocity is
\begin{equation}\label{eq:CellularFlow_b}
 \bm{u}_{\rm e} = \bm{u} - \alpha \epsilon^2  \bm{u} \cdot \nabla \bm{u} \ \ \textrm{with} \ \ \bm{u} = \left(\partial_y \psi,-\partial_x \psi\right) \ \  \textrm{and} \ \ \psi(x,y)=\sin x\sin y,
\end{equation}
where $\alpha = (1-\beta) \gamma = (1-\beta) \St \, \Pe = O(1)$. The inertial contribution turns out to have a gradient structure: we can write \eqref{eq:CellularFlow_b} as
\begin{equation}\label{eq:CellularFlow_c}
 \bm{u}_{\rm e} = \bm{u} + \alpha\epsilon^2\nabla\Phi   \ \ \textrm{with} \ \  \Phi(x,y)=-\left(\cos(2x)+\cos(2y) \right)/4.
\end{equation}
Figure \ref{fig:phidag} displays the $\pi$-periodic potential $\Phi$. In effect, inertia adds a weak potential flow which, depending on the sign of $\alpha$, attracts particles to either the minima or the maxima of $\Phi$ located at the corners or centres of the quarter-cells. 

To compute the effective diffusivity \eqref{eq:Diff_-1}, we first derive the leading-order approximation to $\phi^{\dagger}$, then to $\bm{\varphi}$, and finally combine the resulting expressions. 
\subsection{Asymptotic computation of $\phi^{\dagger}$}

We rewrite Eq.\  \eqref{eq:L0dag} for $\phi^{\dagger}$  as
\begin{equation}\label{eq:CellProbPhidag}
    \epsilon^2 \nabla\phi^{\dagger}-\bm{u}\cdot\nabla\phi^{\dagger}+\alpha\epsilon^2\nabla\cdot\left(\phi^{\dagger}\,\bm{u}\cdot\nabla \bm{u}\right)=0
\end{equation}
and expand  $\phi^{\dagger}$ in powers of $\epsilon^2$ according to $\phi^{\dagger}=\phi^{\dagger}_0+\epsilon^2\phi^{\dagger}_1+\cdots$ to obtain
\begin{subequations}
\begin{align}
    \bm{u}\cdot\nabla\phi^{\dagger}_0&= 0, \label{eq:Phidag_int_order0}\\
    \bm{u}\cdot\nabla\phi^{\dagger}_1&=\nabla^2\phi^{\dagger}_{0}+\alpha\nabla\cdot\left(\phi^{\dagger}_0 \, \bm{u}\cdot\nabla \bm{u}\right)\label{eq:Phidag_int_order1}.
\end{align}
\end{subequations}
Eq.\ \eqref{eq:Phidag_int_order0}  implies that $\phi^{\dagger}_0$ must be constant along streamlines: $\phi^{\dagger}_0=\phi_0^{\dagger}\left(\psi\right)$. Eq.\ \eqref{eq:Phidag_int_order1} can be solved for $\phi^{\dagger}_1$ provided that a solvability condition, obtained by  integrating \eqref{eq:Phidag_int_order1} along streamlines, is satisfied. We show in Appendix \ref{ssec:App:Solvab} that this solvability condition is
\begin{equation}\label{eq:Phidag_int_Solvab}
    \frac{\dd}{\dd\psi}\left(a(\psi)\frac{\dd\phi^{\dagger}_0}{\dd\psi}+\alpha b(\psi)\phi^{\dagger}_0\right)=0,
\end{equation}
where
\begin{equation}\label{eq:PhiDag_int_ab_generic}
a(\psi)=\oint_{\psi}\dd l\, |\bm{u}|  \andand b(\psi)=\oint_{\psi}\dd l\, \bm{n}\cdot\left(\bm{u}\cdot\nabla\bm{u}\right),
\end{equation}
with $l$ denoting arclength and $\bm{n}=\nabla\psi/|\nabla\psi|$ the normal to the streamline. Note that $a(\psi)$ can be recognised as the circulation around each streamline, while $b(\psi)$ is the flux induced by inertial effects through the streamline and can be rewritten as
\begin{equation}
b(\psi) = \oint_\psi\dd l\, k |\bm{u}|^2 = \iint_{D_\psi}\dd \bm{x}\, \det \nabla \nabla \psi  ,
\end{equation}
where $k$ denotes the curvature of the streamline, $D_\psi$ the area enclosed by the streamline and $\nabla \nabla \psi$ the Hessian.
We integrate \eqref{eq:Phidag_int_Solvab} once, taking the integration constant to $0$ as required for a regular solution at the extrema $\psi = \pm 1$. Integrating once more
we find that the (bounded) normalised solution to \eqref{eq:Phidag_int_Solvab} is 
\begin{equation}\label{eq:phidag_FInal}
    \phi^{\dagger}_0\left(\psi\right)=\frac{1}{Z\left(\alpha\right)}\exp\left(-\alpha\int_0^{\psi}\dd \psi'\,\frac{b(\psi')}{a(\psi')}\right),
\end{equation}
where $Z(\alpha)$ is a normalisation constant. After some algebra detailed in appendix \ref{ssec:App:Solvab}, we obtain
\begin{equation}\label{eq:Za_Generic}
    Z(\alpha)=\frac{1}{\pi^2}\int_0^1\dd\psi\,c(\psi)\exp\left(-\alpha\int_0^\psi\dd\psi'\frac{b(\psi')}{a(\psi')}\right),
\end{equation}
where 
\begin{equation}\label{eq:c_def}
    c(\psi)=\oint_\psi\frac{\dd l}{|\bm{u}|}
\end{equation}
is the orbital time around the closed streamlines.

Using the form \eqref{eq:CellularFlow} for the streamfunction we can evaluate the integrals $a(\psi)$, $b(\psi)$ and $c(\psi)$ in terms of (complementary) complete elliptic integrals to find
\begin{subequations}\label{eq:Phidage_int_ab}
\begin{align}
    a(\psi)&=8\left(E'(\psi)-\psi^2 K'(\psi)\right),\\
    b(\psi)&=8\psi\left(K'(\psi)-E'(\psi)\right),\\ 
    c(\psi)&=4 K'(\psi),
\end{align}
\end{subequations}
with $K'(\psi)=K(\sqrt{1-\psi^2})$ and $E'(\psi)=E(\sqrt{1-\psi^2})$ \citep[see][]{DLMF}. {This yields the closed-form}
\begin{equation}\label{eq:Za}
    Z(\alpha)=\frac{4}{\pi^2}\int_{0}^1\dd\psi\,K'(\psi)\exp\left(-\alpha\int_{0}^{\psi}\dd\psi'\,\frac{\psi'\left(K'(\psi')-E'(\psi')\right)}{E'(\psi')-\psi'^2K'(\psi')}\right).
\end{equation}

Figure \ref{fig:phidag} shows the function $\phi^{\dagger}_0$ for three values of $\alpha$ in one quarter-cell. This is of interest because $\phi^{\dagger}$ describes locally the spatial structure of the particle density which is given by $\phi^\dagger(\bm{x})$ times a large-scale diffusive envelope. The function
is even and monotonous with $|\psi|$, decreasing from the centre of quarter-cells for light particles ($\alpha<0$) and increasing from the centre for heavy particles ($\alpha>0$) reflecting the expected clustering induced by inertia. 

\subsection{Asymptotic computation of $\bm{\varphi}$}

It is clear from the symmetries of the cellular flow that  $\bm{c}=0$ and that the effective diffusivity is isotropic. Therefore, we focus on a single component of $\bm{\varphi}$, say $\varphi=\bm{\varphi}\cdot\bm{e}_x$, which satisfies
\begin{equation}\label{eq:varphi_eq}
    \epsilon^2\nabla^2\varphi+\left(\bm{u}-\alpha\epsilon^2 \bm{u}\cdot\nabla \bm{u}\right)\cdot\nabla\varphi+u-\alpha\epsilon^2 \bm{u}\cdot\nabla u=0,
\end{equation}
where $u=\bm{u}\cdot\bm{e}_x$. Making use of the symmetries  $(x,y,\varphi)\mapsto(\pi-y,\pi-x,-\varphi)$, $(x,y,\varphi)\mapsto (x,-y,\varphi)$ and $(x,y,\varphi)\mapsto (-x,y,-\varphi)$ we can focus on the quarter-cell $[0,\pi]^2$ and look for a solution satisfying the boundary conditions
\begin{align}
    \varphi\left(0,y\right)=\varphi\left(\pi,y\right)=0\andand\partial_y\varphi\left(x,0\right)=\partial_y\varphi\left(x,\pi\right)=0.
\end{align}
The general solution of the problem is identical to the solution of this simpler problem up to an irrelevant constant.

We now introduce $\Theta=2(\varphi+x)/\pi-1$ satisfying 
\begin{equation}
    \epsilon^2\nabla^2\Theta+\left(\bm{u}-\alpha\epsilon^2 \left(\bm{u}\cdot\nabla\right)\bm{u}\right)\cdot\nabla\Theta=0,
\end{equation}
with the boundary conditions
\begin{equation}\label{eq:BC_Theta}
    \Theta\left(0,y\right)=-1\comma \Theta\left(\pi,y\right)=1\comma\partial_y\Theta\left(x,0\right)=0\andand\partial_y\Theta\left(x,\pi\right)=0.
\end{equation}
We obtain an approximation for $\Theta$ for $\epsilon \ll 1$ using matched asymptotics.

\subsubsection{Interior solution}
We first consider the solution in the quarter-cell interior. Introducing the expansion $\Theta=\Theta_0+\epsilon^2 \Theta_1+\cdots$, we obtain $\Theta_0=\Theta_0(\psi)$ at leading order and, at the next order,
\begin{equation}\label{eq:varphi_int_order1}
    \bm{u}\cdot\nabla\Theta_1=-\nabla^{2}\Theta_0+\alpha \left( \bm{u}\cdot\nabla \bm{u} \right) \cdot\nabla\Theta_0.
\end{equation}
Integrating along a streamline yields the solvability condition 
\begin{equation}\label{eq:varphi_int_Solvab}
    \frac{\dd}{\dd\psi}\left(a(\psi)\frac{\dd\Theta_0}{\dd\psi}\right)-\alpha b\left(\psi\right)\frac{\dd\Theta_0}{\dd\psi}=0,
\end{equation}
with $a(\psi)$ and $b(\psi)$ given in \eqref{eq:Phidage_int_ab}. The only bounded solution of \eqref{eq:varphi_int_Solvab} is a constant. This constant interior solution must be matched with a boundary layer solution around the separatrix $\psi=0$ so as to satisfy the boundary conditions \eqref{eq:BC_Theta}.

\subsubsection{Boundary-layer solution}

Following \citet{Childress1979}, we introduce the variables
\begin{equation}
    \zeta=\epsilon^{-1}\psi \ \andand \ \sigma=-\int_0^{l}\dd l\,|\bm{u}|,
\end{equation}
where $l$ is the arclength along a streamline. At the separatrix, $0<\sigma<8$ parameterises the boundary of the quarter-cell clockwise, starting from $0$ at  $(0,0)$ and taking values $2$, $4$ and $6$ at successive corners. Introducing the expansion $\Theta=\Theta_0(\sigma,\zeta)+O(\epsilon)$, we obtain at leading order, away from the corners,
\begin{subequations} \label{eq:Soward}
\begin{equation}
    \partial_{\zeta}^2\Theta_0-\partial_{\sigma}\Theta_0=0,
\end{equation}
with the boundary conditions
\begin{align}\label{eq:SowardBC}
    \Theta_0(\sigma,0)=-1 \quad & \text{for}  \ \ \sigma\in[0,2], \quad 
    \Theta_0(\sigma,0)=1 \quad  \text{for} \ \ \sigma\in[4,6],\\
    &\partial_{\zeta}\Theta_0(\sigma,0)=0 \quad \text{for} \ \ \sigma\in[2,4]\cup[6,8].
\end{align}
\end{subequations}
Eqs.\ \eqref{eq:Soward} make up the so-called Childress problem, solved in closed form by \cite{Soward1987}. Using the symmetry of the boundary conditions \eqref{eq:SowardBC}, it can be shown that $\Theta_0$ can only match the constant interior solution if this vanishes. Thus we conclude that, to leading order, $\Theta$ is non-zero only within the boundary layer. As a result, as we now show, we can compute the effective diffusivity by  combining Eq.\ \eqref{eq:phidag_FInal} for $\phi^\dagger$ with the solution of the Childress problem \eqref{eq:Soward}.

\subsection{Computation of the effective diffusivity $\DD_{\rm e}$}

Using the symmetry of the problem, \eqref{eq:hom_varphi_cellprob} and  integration by parts, the effective diffusivity \eqref{eq:Diff_-1} can be recast into the isotropic form 
\begin{equation}\label{eq:Kappa_isotrop}
    \DD_{\rm e}=\overline{D}\,\mathbb{I} \quad \textrm{with} \quad  \overline{D}=\epsilon^2\left\langle\phi^{\dagger}\left(2\partial_{x}\varphi+|\nabla\varphi|^2+1\right)\right\rangle.
\end{equation}
Since the four quarter-cells are equivalent,  we can focus on $[0,\pi]^2$ where $\varphi=\pi(1+\Theta)/2-x$ to obtain
\begin{equation}
    \overline{D}=\frac{\pi^2\epsilon^2}{4}\left\langle\phi^{\dagger}|\nabla\Theta|^2\right\rangle_{[0,\pi]^2}
\end{equation}
where $\langle\cdot\rangle_{[0,\pi]^2}$ denotes the spatial average over the quarter-cell $[0,\pi]^2$. In the limit $\epsilon\ll1$, $\Theta$ becomes a boundary layer term localised in the $O(\epsilon)$ boundary layer around the separatrix  $\psi=0$. Within this boundary layer, $\phi^{\dagger}\sim Z(\alpha)^{-1}$ as \eqref{eq:phidag_FInal} shows. As a result, to leading order in $\epsilon$, the effective diffusivity reduces to
\begin{equation}\label{eq:kappa_int}
    \overline{D} \sim \frac{\epsilon}{4Z(\alpha)}\int_{0}^8\dd\sigma\int_0^{\infty}\dd\zeta\left(\partial_{\zeta}\Theta_0\right)^2
\end{equation}
using that $\dd x \dd y = \epsilon^{-1}|\nabla \psi|^{-2}\dd \sigma \dd \zeta$.
In Appendix \ref{ssec:app:intcomp}, we relate the integral in \eqref{eq:kappa_int} to Soward's constant ${\So}$ appearing in the asymptotic expression of the non-inertial effective diffusivity \eqref{eq:Diff_passive} to rewrite the effective diffusivity in the fully explicit form \eqref{eq:Effective_Diff_MainResult}.
Note that it is possible to bypass to the computation in Appendix \ref{ssec:app:intcomp} by observing that the integral in \eqref{eq:kappa_int} is independent of $\alpha$ so that it can be determined by setting $\alpha=0$  in \eqref{eq:kappa_int}  and matching to the familiar non-inertial result \eqref{eq:Diff_passive}, noting that $Z(0)=1$.

\section{Discussion and conclusion\label{sec:5}}

\begin{figure}
    \centering
    \includegraphics[width=\linewidth]{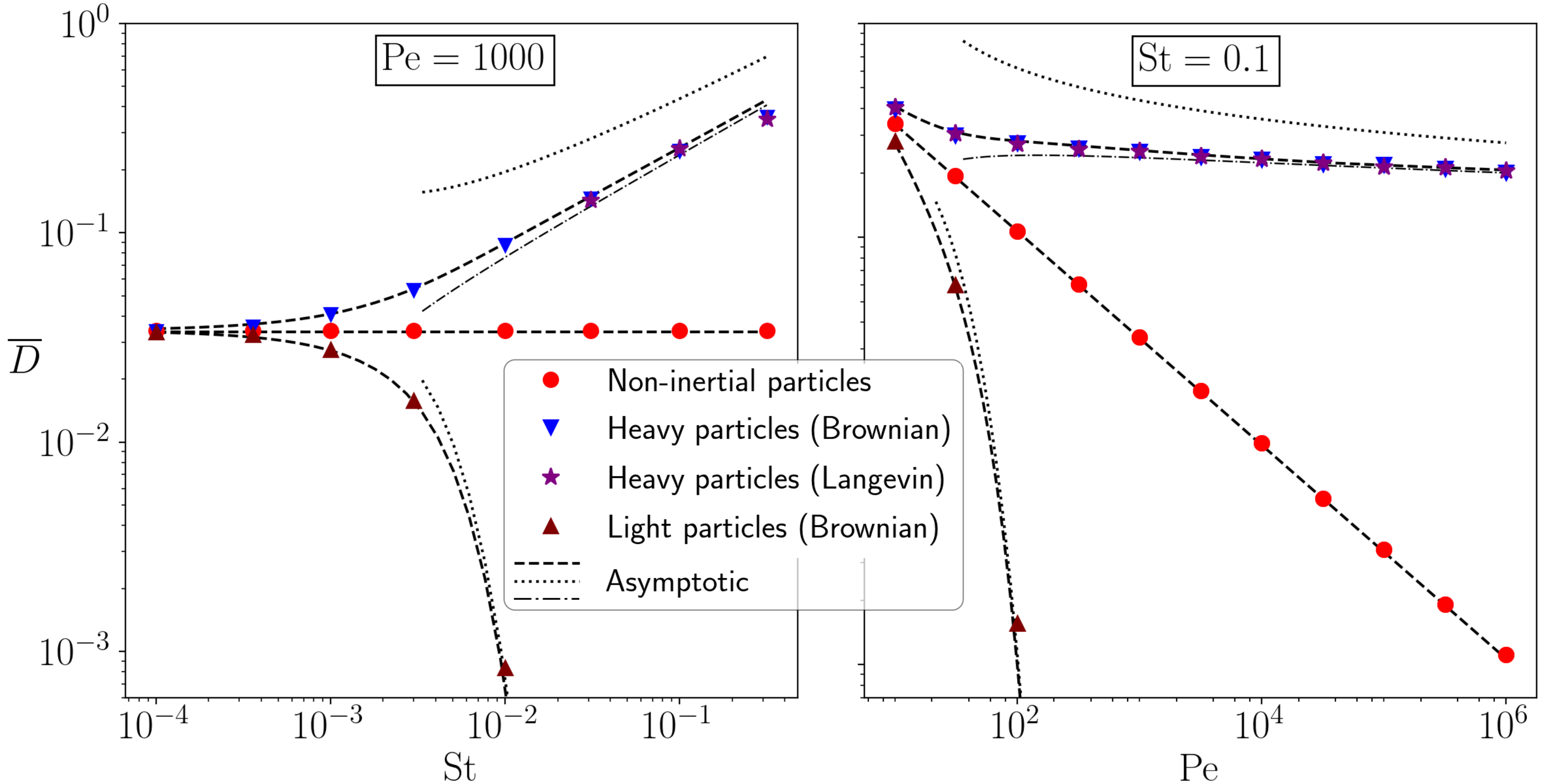}
    \caption{Effective diffusivity $\overline{D}$ as a function of $\St$ for $\Pe=1000$ (left) and as a function of $\Pe$ for $\St=0.1$ (right). The values estimated by direct numerical sampling of the Langevin dynamics \eqref{eq:Langevin} for $\beta=0$ ($\star$), the effective Brownian dynamics \eqref{eq:Overdamped-Langevin} for $\beta=0$ ($\bigtriangledown$) and $\beta=2$ ($\bigtriangleup$), and  the non-inertial dynamics \eqref{eq:PassivScal_Langevin} ($\circ$) are compared to the asymptotic prediction  \eqref{eq:Effective_Diff_MainResult}. The simplified asymptotic expressions 
    \eqref{eq:EffDiff_aG0} and \eqref{eq:EffDiff_aM0} valid for $\St \, \Pe \gg 1$ are shown (dotted lines), as well as the refinement of \eqref{eq:EffDiff_aG0} that includes the logarithmic correction in \eqref{eq:asympZa} (dot-dashed line).}
    \label{fig:Asymmp}
\end{figure}

The key result of the paper is the asymptotic expression  \eqref{eq:Effective_Diff_MainResult} for the effective diffusivity $\overline{D}$ in the distinguished regime $\St = O(\Pe^{-1}) \ll1$. We test it against the direct numerical sampling of the dynamics of the particles. The  Langevin equation  \eqref{eq:Langevin} can be costly for $\St \ll 1$ so we restrict its use to moderately small $\St$; for smaller $\St$, we instead integrate the effective Brownian dynamics  \eqref{eq:Overdamped-Langevin}. We use numerical schemes based on those of \cite{Pavliotis2009} {which split the flow over one timestep between two advection steps (each carried out exactly using the variables $x+y$ and $x-y$) and a diffusion step. The exact area-preservation of the advection steps is essential to enable large time steps and long integration times.} The results are summarised in Figure \ref{fig:Asymmp}. They demonstrate a very good agreement between direct estimations and asymptotic predictions of $\overline{D}$. The near-coincidence of estimates obtained with the Langevin and effective Brownian dynamics also confirms the validity of the latter.

The impact of inertia is entirely captured by the function $Z(\alpha)$ defined by \eqref{eq:Za} and shown in Figure \ref{fig:Za}.  We now analyse the behaviour of this function in detail. As noted in \S1, $Z(\alpha)$ is decreasing with $\alpha$, with $Z(\alpha)<1$  for $\alpha<0$ and $Z(\alpha)>1$ for $\alpha>0$. Consequently, the effective diffusivity of particles denser (resp.\ less dense) than the fluid is always larger (resp.\ smaller) than that of non-inertial particles. This can be attributed to the divergence  \eqref{eq:DivU} of the effective velocity which, when averaged along streamlines, leads to an accumulation (resp.\ depletion) of particles in the separatrix region (cf.\ Figure \ref{fig:phidag}), which controls the cell-to-cell and hence global transport.

It is interesting to consider the limiting behaviour of $Z(\alpha)$ and hence $\overline{D}$ as $\alpha = \St \, \Pe (1-\beta) \to \pm \infty$, that is, when inertia dominates over diffusion. 
In the heavy-particle case $\alpha \to \infty$, the asymptotics of $Z(\alpha)$ derived in Appendix \ref{ssec:App:asympZ} gives
\begin{equation}\label{eq:EffDiff_aG0}
    \overline{D} \sim {\So}\pi^{3/2}\left(\frac{\St(1-\beta)}{\log \left(\Pe \, \St (1-\beta) \right)}\right)^{1/2} \quad \textrm{for} \ \ \Pe^{-1}\ll\St\ll1 \ \ \textrm{and} \ \ \beta < 1.
\end{equation}
This corresponds to an effective diffusivity that depends only weakly on the P\'eclet number or, equivalently on molecular diffusivity, and is instead controlled by inertia through the dependence on $\St^{1/2}$. Note that the asymptotics \eqref{eq:EffDiff_aG0} is rather poor because it ignores logarithmic corrections that are negligible only for exceedingly large $\alpha$. A more accurate formula can be obtained by using an improved approximation to $Z(\alpha)$ given in \eqref{eq:asympZa} and shown in Figure \ref{fig:Za}. The predictions of \eqref{eq:EffDiff_aG0} and its improvement are compared against simulations results and the full asymptotic approximation \eqref{eq:Effective_Diff_MainResult} in Figure \ref{fig:Asymmp}.

In the light-particle case $\alpha \to - \infty$, the asymptotics of $Z(\alpha)$ in Appendix \ref{ssec:App:asympZ} gives 
\begin{equation}\label{eq:EffDiff_aM0}
    \overline{D} \sim {\So}\pi\Pe^{1/2}\St(\beta-1)\,\textrm{e}^{-\Upsilon \, \St\, \Pe(\beta-1)} \quad \textrm{for} \ \ \Pe^{-1}\ll\St\ll1 \ \ \textrm{and} \ \ \beta > 1,
\end{equation}
where $\Upsilon\approx0.655$, which is also shown in Figure \ref{fig:Asymmp}. The effective diffusivity decreases exponentially with $\Pe \, \St$ corresponding to a dramatic inhibition of dispersion caused by inertia. Physically, particles are trapped by inertia near the quarter-cell centres and only escape by crossing the separatrix for rare realisations of the noise. The rate of these rare escapes and hence $\overline{D}$ could be estimated using small-noise large-deviation theory \citep[e.g.][]{Freidlin12}. Note that the key parameter $\St \, \Pe = U^2 \tau/D$ (which is independent from the flow scale $a$) can be rewritten as the Arrhenius-like number
\begin{equation} \label{eq:arrhenius}
    \St \, \Pe = {m U^2}/(k_B T), 
\end{equation}
where $m=m_\mathrm{p}+m_\mathrm{f}/2$ is the effective mass of the particles, $k_\mathrm{B}$ is the Boltzmann constant and $T$ the temperature, on using the Einstein--Smoluchowski relation $D = \tau k_\mathrm{B} T / m_{\rm p}$. Thus \eqref{eq:EffDiff_aM0} can be interpreted as a form of Arrhenius law, with the particle kinetic energy playing the role of activation energy.

We can assess the relative importance of inertia and diffusion by thermal noise for particles in a flow using \eqref{eq:arrhenius}. For instance,  at room temperature, for particles with an effective density similar to that of water, we find that $\St\,\Pe \approx r_\mathrm{p}^3 U^2 \times 10^{24}$ in SI units. Thus, inertia dominates  diffusion ($\St\,\Pe \gtrsim 1$) for particles of radius $r_\mathrm{p}=10^{-6}$ m and $10^{-3}$ m when $U \gtrsim 10^{-3}$ m/s and $U \gtrsim 10^{-8}$ m/s, respectively. Since typical flow velocities are likely to exceed these small values, the simplified asymptotic expressions \eqref{eq:EffDiff_aG0}--\eqref{eq:EffDiff_aM0} will be valuable. We emphasise that, though subdominant, diffusion plays a crucial role in setting the effective diffusivity of cellular flows, as the dependence in $\Pe$ of \eqref{eq:EffDiff_aG0}--\eqref{eq:EffDiff_aM0} indicates. This is because diffusion is indispensible for particles to move between cells. We also note that, while \eqref{eq:arrhenius} is restricted to diffusion by thermal noise, our results are also relevant to dispersion problems where diffusion models small-scale turbulent mixing, in which case the (turbulent) diffusivity is orders of magnitude larger than the molecular one and $\St \, \Pe$ is not necessarily large.

We remark that the $\Pe^{-1/2}$ scaling for the effective diffusivity of non-inertial particles in \eqref{eq:Diff_passive} is a universal feature of periodic flows with closed streamlines \citep{Heinze2003,Novikov2005}. We expect that the conclusion that this is corrected to $\Pe^{-1/2}/Z(\alpha)$ when inertia is taken into account, which we draw for the cellular flow \eqref{eq:CellularFlow}, also generalises. {Specifically, we expect that
\begin{equation}
    \overline{D}(\Pe,\alpha) = {\overline{D}(\Pe,0)}/Z(\alpha)
    \label{eq:DDZ}
\end{equation}
for all cellular flows.}
The form of $Z(\alpha)$ is specific to each flow but its qualitative dependence on $\alpha$ and its asymptotic scalings for $\alpha \to \pm \infty$ are likely to be as in the case of the {canonical} cellular flow \eqref{eq:CellularFlow}. {To illustrate this, we consider another cellular flow in appendix \ref{sec:App:NewFlow} and confirm \eqref{eq:DDZ}.} Flows with open streamlines behave very differently, however. In shear flows, numerical results from  \cite{Pavliotis2006} suggests that inertia only has a negligible effect on the (Taylor) effective diffusivity. It would be of interest to examine the impact of inertia on more complex flows such as the cat's-eye flows of  \citet{Childress1989}. 

{It would also be desirable to assess the effect of the Boussinesq--Basset force which we neglected. Although \cite{Manton1974} argues that it is less significant than other inertial effects, more recent work by \cite{Daitche2011} and \cite{Langlois2015} suggests otherwise.}

\medskip

\noindent
\textbf{Acknowledgments.} This work was supported by EPSRC Programme Grant EP/R045046/1: Probing Multiscale Complex Multiphase Flows with Positrons for Engineering and Biomedical Applications (PI: Prof.\ M. Barigou, University of Birmingham). {We thank the two anonymous referees for useful comments.}\\
\textbf{Declaration of Interests.} The authors report no conflict of interest.

\appendix
\section{Derivation of the solvability condition \eqref{eq:Phidag_int_Solvab}\label{ssec:App:Solvab}}
We introduce the time-like coordinate $s$ such that
\begin{equation}\label{eq:timelike}
    \frac{\dd}{\dd s}=\bm{u}\cdot\nabla.
\end{equation}
Integrating \eqref{eq:Phidag_int_order1} along streamlines gives
\begin{equation}\label{eq:App_1_Solva}
\oint_{\psi}\dd s\,\nabla\cdot\left(\nabla\phi_0+\alpha\phi_0\, \bm{u}\cdot\nabla\bm{u}\right)=0.
\end{equation}
Following \cite{HaynesVanneste2014}, we use the fact that the change of variables $(x,y)\mapsto (s,\psi)$ is area preserving and the divergence theorem to write
\begin{equation}
    \oint_{\psi}\dd s\, \nabla\cdot \bm{f}=\frac{\dd}{\dd\psi}\left(\iint \dd x\dd y\, \nabla\cdot\bm{f}\right)=\frac{\dd}{\dd\psi}\left(\oint_{\psi}\dd s\, \nabla \psi\cdot \bm{f}\right)
\end{equation}
for an arbitrary vector field $\bm{f}$. This reduces \eqref{eq:App_1_Solva} to
\begin{equation}
    \frac{\dd}{\dd\psi}\left(a(\psi)\frac{\dd\phi_0}{\dd\psi}+\alpha b(\psi)\phi_0\right)=0,
\end{equation}
where
\begin{equation}
    a(\psi)=\oint_{\psi}\dd s\, |\nabla\psi|^2\quad \textrm{and} \quad  b(\psi)=\oint_{\psi}\dd s\,\nabla\psi\cdot\left(\bm{u}\cdot\nabla\bm{u}\right).
\end{equation}
{Using that $\dd s=|\nabla\psi| \, \dd l$ with $l$ the arclength, we obtain \eqref{eq:PhiDag_int_ab_generic}.}
The derivation of \eqref{eq:Phidage_int_ab} for $a(\psi)$ {from \eqref{eq:CellularFlow}} can be found in \cite{HaynesVanneste2014}. To compute $b(\psi)$, we use the symmetry of the streamline,  \eqref{eq:timelike} and $\psi= \sin x\sin y$, to write
\begin{equation}
    b(\psi)=\psi\int_{\sin^{-1}\psi}^{\pi/2}\frac{\cos^2 x \, \dd x}{\sqrt{\sin^{2}x-\psi^2}}.
\end{equation}
The substitution $t^2 = (1-\sin^2x)/(1-\psi^2)$ then gives
\begin{equation}
    b(\psi)=\psi(1-\psi^2)\int_{0}^{1}\frac{t^2 \, \dd t}{\sqrt{1-t^2}\sqrt{1-(1-\psi^2)t^2}}.
\end{equation}
Using formula (19.2.6) in \cite{DLMF}, this reduces to the expression in \eqref{eq:Phidage_int_ab}.

\section{Computation of the integral in \eqref{eq:kappa_int}\label{ssec:app:intcomp}}
We show that the integral 
\begin{equation}\label{eq:app:IntK}
    \mathcal{K}=\int_0^8\dd\sigma\int_0^{\infty}\dd\zeta\,\left(\partial_{\zeta}\Theta_0\right)^2,
\end{equation}
where $\Theta_0$ solves \eqref{eq:Soward}, is related to
\begin{equation}\label{eq:SowardResult}
    \int_{0}^{\infty}\dd\zeta\,\Theta_0\left(2,\zeta\right)=-2{\So} \ \with \ {\So}=\sqrt{\frac{2}{\pi}}\sum_{n=0}^{\infty}\frac{(-1)^n}{\sqrt{2n+1}},
\end{equation}
as calculated by  \cite{Soward1987}. Integrating \eqref{eq:app:IntK} by parts in $\zeta$ gives
\begin{equation}\label{eq:K_Step_1}
    \mathcal{K}=-\int_{0}^8\dd\sigma\,\Theta_0(\sigma,0)\partial_{\zeta}\Theta_0(\sigma,0)-\int_{0}^{8}\dd\sigma\int_0^{\infty}\dd\zeta\,\Theta_0\partial_{\zeta}^2\Theta_0.
\end{equation}
The second term can be shown to vanish using \eqref{eq:Soward} and periodicity in $\sigma$. Using the boundary condition \eqref{eq:SowardBC} reduces the first term to
\begin{equation} \label{eq:aaa}
    \mathcal{K}=\int_0^2\dd\sigma\,\left(\partial_{\zeta}\Theta_0(\sigma,0)-\partial_{\zeta}\Theta_0(\sigma+4,0)\right).
\end{equation}
Now, integrating \eqref{eq:Soward} for $\zeta\in[0,\infty)$ gives  $\int_{0}^{\infty}\dd\zeta\,\partial_{\sigma}\Theta_0=-\partial_{\zeta}\Theta_0(\sigma,0)$ which can be introduced in \eqref{eq:aaa} to obtain
\begin{equation}
    \mathcal{K}=\int_{0}^{\infty}\dd\zeta\,\left(\Theta_{0}(0,\zeta)-\Theta_{0}(2,\zeta)+\Theta_{0}(6,\zeta)-\Theta_{0}(4,\zeta)\right).
\end{equation}
Finally, using that $\int_{0}^{\infty}\dd\zeta\,\Theta_0(\sigma,\zeta)$ is constant for $\sigma \in [2,4]$ and $[4,8]$ and the symmetry $\Theta_0(\sigma+4,\zeta)=-\Theta_0(\sigma,\zeta)$, we obtain 
\begin{equation}
    \mathcal{K}=-4\int_0^{\infty}\dd\zeta\,\Theta_0\left(2,\sigma\right)=8{\So}.
\end{equation}

\section{Asymptotic calculation of $Z(\alpha)$ for $\alpha\to\pm\infty$ \label{ssec:App:asympZ}}
\subsection{Limit $\alpha \to \infty$}

The outer integral in \eqref{eq:Za} is dominated by a neighbourhood of the minimum of the integral multiplying $\alpha$. This minimum can be verified to be the left endpoint $\psi=0$.  Using asymptotic expansions for $K'(\psi)$ and $E'(\psi)$ as $\psi\to 0^+$, we find the approximation
\begin{equation}
    Z\left(\alpha\right) \sim \frac{4}{\pi^2}\int_0^1\dd\psi\,\left(-\log\frac{\psi}{4}\right)\exp\left(\alpha\frac{\psi^2}{4}\left(1+2\log\frac{\psi}{4}\right)\right).
\end{equation}
Introducing the new integration variable $x = -\alpha\psi^2(1+2\log(\psi/4))/4$ yields, after cumbersome calculations (carried out using the symbolic-algebra software Mathematica), 
\begin{equation}
\label{eq:messyInt}
Z\left(\alpha\right) \sim \frac{2}{\pi^2\sqrt{\alpha}}\int_0^{\alpha c} \frac{(w(x)+1)\sqrt{w(x)} \mathrm{e}^{-x}}{w(x)-1}\,\dd x,
\end{equation}
where $c=\log 2-1/4>0$ and $w(x)=-W_{-1}\left(-\mathrm{e}\,x/(4\alpha)\right)$, with
$W_{-1}$ the $-1$-branch of the Lambert-$W$ function, namely the inverse function of $x\mathrm{e}^{x}$ on $(-\infty,-1]$ \citep[see][]{DLMF}. Using the asymptotic expansion
\begin{equation}
    W_{-1}(-x) = \log x+\log(-\log x)+\frac{\log(-\log x)}{\log x}+o\left(\frac{\log(\log x)}{\log x}\right) \ \ \textrm{as} \ \ x \to 0^+, 
\end{equation}
we expand the integrand in \eqref{eq:messyInt} for $\alpha\to \infty$ then extend the integration interval to $[0,\infty)$ to find
\begin{equation}\label{eq:asympZb}
    Z\left(\alpha\right) \sim \frac{2\sqrt{\log\alpha}}{\pi^2\sqrt{\alpha}}\left(\left(1+\frac{3+2\log2+\log \log\alpha}{2\log\alpha}\right)I_{0}+\frac{1}{2\log\alpha}I_{1}\right),
\end{equation}
where 
\begin{equation}\label{eq:defIn}
I_{0}=\int_{0}^{\infty}\frac{\mathrm{e}^{-x} \, \dd x}{\sqrt{x}}=\sqrt{\pi}\ \andand \ I_{1}=-\int_0^{\infty}\frac{\mathrm{e}^{-x}\log x \, \dd x}{\sqrt{x}}=\sqrt{\pi}\left(\gamma_\mathrm{E}+2\log 2\right)
\end{equation}
with  $\gamma_\mathrm{E} =  0.577\cdots$ the  Euler--Mascheroni constant.
Combining \eqref{eq:asympZb} with \eqref{eq:defIn} finally yields
\begin{equation}\label{eq:asympZa}
    Z\left(\alpha\right) \sim \frac{2\sqrt{\log\alpha}}{\pi^{3/2}\sqrt{\alpha}}\left(1+\frac{3+\gamma_\mathrm{E}+\log(16\log\alpha)}{2\log\alpha}\right) \quad \textrm{as} \ \ \alpha \to \infty.
\end{equation}
 Note that the corrections up to $(\log\alpha)^{-1}$ are necessary for numerical applications as they decay very slowly with $\alpha$ (see Figure \ref{fig:Za}).

\subsection{Limit $\alpha \to -\infty$}

In this case, the outer integral in \eqref{eq:Za} is dominated by a neighbourhood of the maximum of the integral multiplying $\alpha$, located at the right endpoint $\psi=1$. Expanding the inner integral near $\psi=1$  we find
\begin{equation}\label{eq:Za_Arheynius}
    Z(\alpha) \sim \frac{2}{\pi|\alpha|}\mathrm{e}^{\Upsilon|\alpha|} \quad \textrm{as} \ \ \alpha \to - \infty, \quad \textrm{where} \ \  \Upsilon=\int_{0}^{1}\frac{b(\psi)}{a(\psi)} \, \dd\psi = 0.655\cdots
\end{equation}
using that  $K'(1)=\pi/2$ and $b(\psi)/a(\psi)\to 1$ as $\psi \to 1$.

\section{Another cellular flow\label{sec:App:NewFlow}}
\begin{figure}
    \centering
    \includegraphics[width=\linewidth]{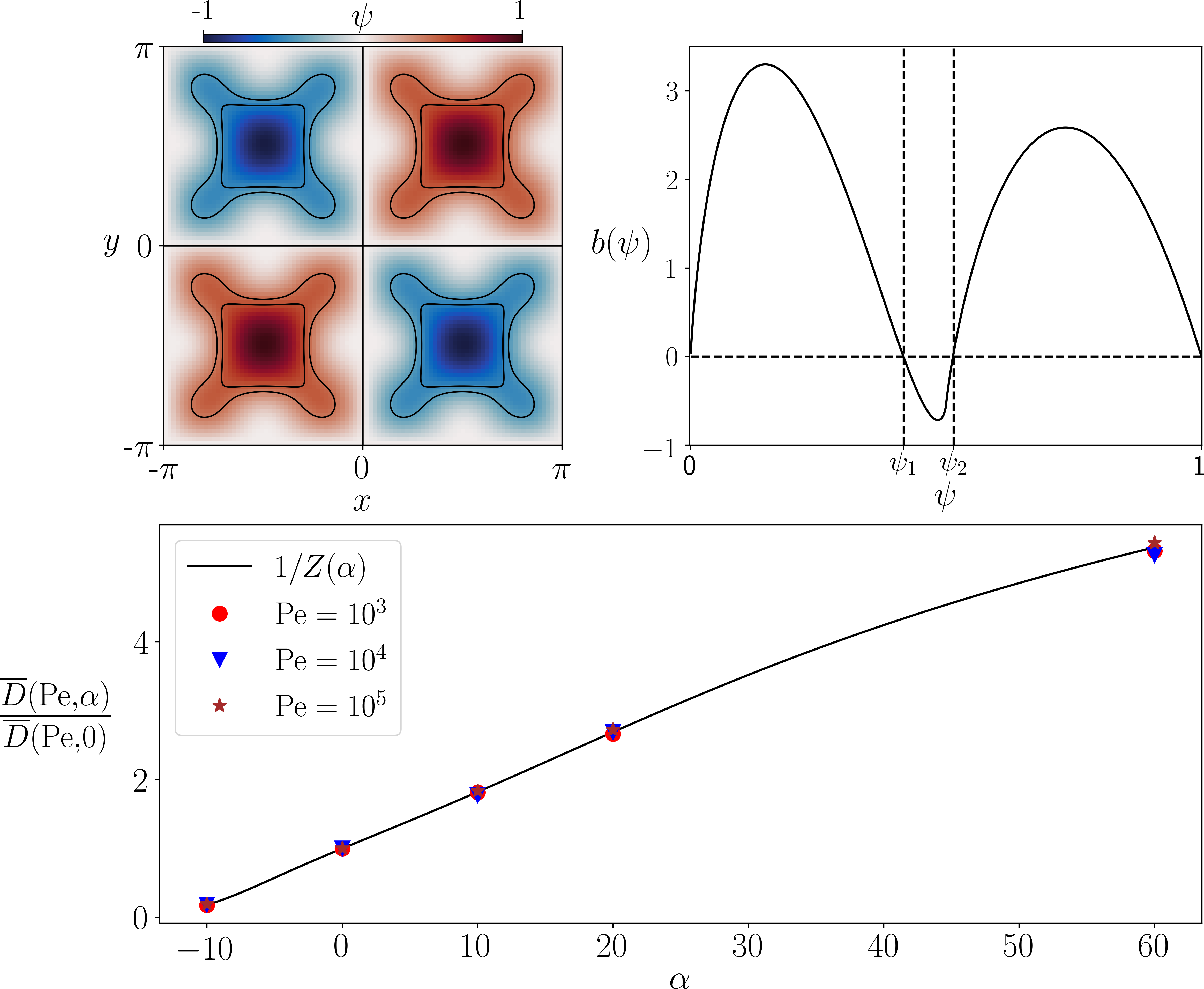}
    \caption{ Top-left: streamfunction \eqref{eq:NewFLow}.  Top-right: function $b(\psi)$ for the streamfunction  \eqref{eq:NewFLow} computed from \eqref{eq:PhiDag_int_ab_generic}. The streamlines associated with the roots $\psi_1\approx 0.42$ and $\psi_2\approx 0.52$ are represented with solid lines in the top-left panel. {Bottom:} ratio of inertial over non-inertial particles effective diffusivities $\overline{D}$ as a function of $\alpha$ for $\Pe=10^3$ (circles), $10^4$ (triangles) and $10^6$ (stars) (right panel). The effective diffusivities are estimated by direct numerical sampling of the effective Brownian dynamics \eqref{eq:Overdamped-Langevin}.  Results are compared against a numerical  evaluation of  $1/Z(\alpha)$ from \eqref{eq:Za_Generic}.}
    \label{fig:Newflow}
\end{figure}
In this appendix, we examine a cellular flow different from the canonical cellular flow \eqref{eq:CellularFlow} in order to verify \eqref{eq:DDZ}.
We consider the flow with streamfunction
\begin{equation}\label{eq:NewFLow}
    \psi(x,y)=\tfrac{3}{4}\sin x\sin y+\tfrac{1}{4}\sin (3x)\sin (3y).
\end{equation}
A period of this flow, displaying the streamfunction and  corresponding velocity field, is shown in the top-left panel of figure \ref{fig:Newflow}.

In the bottom panel of figure \ref{fig:Newflow}, we compare direct numerical estimates of the effective diffusivity ratio  $\overline{D}(\Pe,\alpha)/\overline{D}(\Pe,0)$ to the value  of $1/Z(\alpha)$ computed from \eqref{eq:Za_Generic}, where the integrals along streamlines $a(\psi)$, $b(\psi)$ and $c(\psi)$ are evaluated from their definitions \eqref{eq:PhiDag_int_ab_generic} and \eqref{eq:c_def}. The good match confirms the validity of \eqref{eq:DDZ}.

We also note here that for \eqref{eq:NewFLow} the function  $b(\psi)$ becomes negative for a range of streamlines inside the quarter-cells. Within the quarter-cells with counter-clockwise rotation, these streamlines are characterised by $0<\psi_1\leq \psi\leq\psi_2<1$ (see top-right panel of figure  \ref{fig:Newflow}). This is significant because $b(\psi)$ represents a (suitably averaged) cross-streamline drift induced by the weak inertia, with $\alpha b(\psi)>0$ corresponding to drift towards the separatrix, and $\alpha b(\psi)<0$ towards the centres of the quarter-cells. Therefore, in the absence of diffusion particles heavier than fluid ($\alpha>0$) initially on streamlines $0<\psi<\psi_1$ are attracted by the separatrix $\psi=0$ while particles on streamlines $\psi_1 < \psi < 1$ are attracted by the streamline $\psi=\psi_2$. 
In the presence of diffusion, the overall effect remains an expulsion of the particles from the cells towards the separatrices leading to enhanced dispersion because $Z(\alpha)<1$, but this is only established on long time scales due to the presence of the attracting streamlines $\psi=\psi_2$. 

The change of sign of $b(\psi)$ raises the intriguing possibility that for certain flows inner streamlines can be sufficiently attractive that $Z(\alpha)>1$ for some $\alpha>0$, so that heavy particles concentrate in the interior of the cells and dispersion is inhibited by inertia. We have not been able to find such flows nor rule out their existence.

\bibliographystyle{jfm}
\bibliography{Reference}

\end{document}